\documentclass[preprint,
amsmath,amssymb,amsfonts,aps,prd,superscriptaddress,nofootinbib]{revtex4-1}
\usepackage{ifsym}
\usepackage{MnSymbol}
\usepackage[dvipsnames]{xcolor}
\usepackage{hyperref} 
\hypersetup{
    pdftex,
    pdfstartview=FitH,
    bookmarksnumbered=true,
    pagebackref,
    colorlinks,
    breaklinks,
    urlcolor=Maroon,
    linkcolor=Maroon,
    citecolor=Bittersweet
}
\usepackage{tikz}
\usetikzlibrary{calc,matrix,decorations.pathmorphing,decorations.markings,arrows,positioning,intersections,mindmap,backgrounds}

\def\be{\begin{equation}}
\def\ee{\end{equation}}
\def\ba{\begin{eqnarray}}
\def\ea{\end{eqnarray}}
\def\nl{\nonumber\\}

\def\CP1{\mathbb{CP}^1}
\def\SL2C{\mathrm{SL}(2,\mathbb{C})}

\def\Z2{\mathbb{Z}_2}

\def\su2{{SU(2)}}

\def\[{\left[}
\def\]{\right]}

\def\({\left(}
\def\){\right)}
\def\[{\left[}
\def\]{\right]}

\def\<{\langle}
\def\>{\rangle}

\def\i2{\frac{i}{2}}

\def\2F1{\,_2{\rm F}_1}

\newcommand{\beq}{\begin{equation}}
\newcommand{\eeq}{\end{equation}}
\newcommand{\beqq}{\begin{equation*}}
\newcommand{\eeqq}{\end{equation*}}
\newcommand\beqa{\begin{eqnarray}}
\newcommand\eeqa{\end{eqnarray}}
\newcommand\beqaa{\begin{eqnarray*}}
\newcommand\eeqaa{\end{eqnarray*}}
\newcommand\bea{\begin{array}}
\newcommand\eea{\end{array}}

\begin{document}


\title{One-loop Scattering Equations and Amplitudes\\ from Forward Limit}


\author{Song He}
\email{she@perimeterinstitute.ca}
\affiliation{Perimeter Institute for Theoretical Physics, Waterloo, ON N2L 2Y5, Canada}
\affiliation{School of Natural Sciences, Institute for Advanced Study, Princeton, NJ 08540, USA}
\author{Ellis Ye Yuan}
\email{yyuan@perimeterinstitute.ca}
\affiliation{Perimeter Institute for Theoretical Physics, Waterloo, ON N2L 2Y5, Canada}
\affiliation{Department of Physics \& Astronomy, University of Waterloo, Waterloo, ON N2L 3G1, Canada}


\date{\today}

\begin{abstract}
We show that the forward limit of tree-level scattering equations with two massive particles yields the $\SL2C$-covariant form of the one-loop scattering equations  recently proposed by Geyer {\it et al}. We clarify several properties about these equations and the formulas at one loop. We then argue that in the bi-adjoint scalar theory, such forward limit yields the correct one-loop massless amplitudes, which leads to a new formula for the latter. 
\end{abstract} 


\maketitle 

\section{Introduction}

For a large variety of massless theories in arbitrary dimensions, the tree-level S-matrix can be compactly formulated as an integral over the moduli space of Riemann spheres~\cite{Cachazo:2013hca, Cachazo:2013iea,Cachazo:2014nsa,Cachazo:2014xea}. The key ingredient in this construction is the tree-level scattering equations~\cite{Cachazo:2013gna}
\be
\sum_{b\neq a} \frac{k_a\cdot k_b}{\sigma_{a}-\sigma_{b}} = 0\,, \quad a\in \{ 1,2,\ldots , n\}\,,
\label{scatt}\ee
where $\sigma_a$ denotes the position of the $a^{\rm th}$ puncture. These equations have made an appearance in previous literature in different contexts \cite{Fairlie:1972, *Roberts:1972, *Fairlie:2008dg, Gross:1987ar, Witten:2004cp, Makeenko:2011dm,Cachazo:2012uq}. The tree-level S-matrix admits a representation (sometimes called CHY formula or CHY representation) of the form
\be\label{generalformula}  
{\cal M}^{\rm tree}_n = \int \frac{d\,^n\sigma}{\textrm{vol}\,\SL2C} \prod_a {}'\delta(\sum_{b\neq a} \frac{k_a\cdot k_b}{\sigma_{a}-\sigma_{b}})~{\cal I}_n(\{k,\epsilon,
\sigma\})\,,
\ee
where the ${\cal I}_n$ depends on the theory and carries all the information about wave functions of external particles (such as polarizations $\epsilon$'s). This formula is invariant under $\SL2C$ transformations, and one has to fix three punctures and remove three delta constraints (as indicated by the notation $\prod_a'$) to get rid of this redundancy. For some of the more recent works on scattering equations and the formula, see~\cite{Dolan:2013isa,Schwab:2014xua,Afkhami-Jeddi:2014fia,Zlotnikov:2014sva,Kalousios:2014uva,Cachazo:2015ksa,Cachazo:2015nwa,Baadsgaard:2015voa,Baadsgaard:2015ifa}.

Elegant worldsheet models known as ambitwistor string theory have been argued to underpin scattering equations and CHY formulas for tree amplitudes in various theories~\cite{Mason:2013sva,Berkovits:2013xba} (see~\cite{Adamo:2014wea,Ohmori:2015sha,Casali:2015vta,Adamo:2015gia} for recent progress). In ambitwistor string theory the formalism can be extended to Riemann surfaces of higher genus, as first studied by Adamo, Casali and Skinner in~\cite{Adamo:2013tsa}. They proposed scattering equations on a torus and conjectured a formula for one-loop amplitudes in supergravity in ten dimensions, which were further studied in~\cite{Adamo:2013tca,Casali:2014hfa}.

Based on these developments, Geyer, Mason, Monteiro and Tourkine proposed novel formulas for one-loop amplitudes in maximal supergravity and super Yang--Mills theories~\cite{Geyer:2015bja}. The key ingredient therein is a set of $n-1$ equations, dubbed as off-shell scattering equations, derived from the torus scattering equations by a residue theorem~\cite{Geyer:2015bja}:
\be\label{offsca}
\sum_{b=1, b\neq a}^n  \frac{k_a \cdot k_b}{\sigma_{a}-\sigma_{b}}+\frac{k_a\cdot \ell}{\sigma_a}=0\,,\quad a\in \{2,\ldots , n\}\,,
\ee
where the off-shell momentum $\ell$ ($\ell^2\neq 0$) is interpreted as the loop momentum, and the $\sigma$'s are again treated as punctures on a Riemann sphere. The equation for $a=1$ is absent and $\sigma_1$ is fixed, which is reminiscent of the translation redundancy on the torus. The formula is much simpler than that in~\cite{Adamo:2013tsa}, rather it takes a similar form as the tree-level CHY formula~\eqref{generalformula}.

It is thus natural to ask: is it possible to obtain the off-shell equations and formulas for one-loop amplitudes without knowing the ambitwistor strings? Would this indicate any new connections between one-loop amplitudes and tree-level ones?

In this paper, we argue that the key to these questions is to consider tree-level scattering of massless particles with two additional {\it massive} particles that probe the forward limit \cite{Feynman:1963ax,CaronHuot:2010zt}. The limit of tree-level equations yields the $\SL2C$-covariant form of \eqref{offsca}, which we call one-loop scattering equations (not to confuse with those of~\cite{Adamo:2013tsa}). Scattering equations with massive particles have been studied in~\cite{Dolan:2014ega,Naculich:2014naa,Naculich:2015zha,Naculich:2015coa}. In the particular case of $n$ massless particles and two massive ones, as shown by Naculich~\cite{Naculich:2014naa,Naculich:2015zha}, the equations are only slightly modified and a tree amplitude can be written in the CHY formulation as~\eqref{generalformula}. We find the forward limit to be non-trivial: out of the $(n-1)!$ solutions to the massive scattering equations, $(n-2)!$ drop out, while the remaining $(n-1)!-(n-2)!$ become the solutions to the one-loop equations. In comparison, \eqref{offsca}, which naturally arise from ambitwsitor strings, only pick out $(n-1)!-2(n-2)!$ among these solutions. This does not lead to any contradiction for the formulas considered in \cite{Geyer:2015bja}, since the missing $(n-2)!$ solutions do not contribute in those special cases. However, in general cases all the $(n-1)!-(n-2)!$ solutions contribute.

For general theories it is not completely clear whether the forward limit of an amplitude with two massive particles prescribed above makes sense as certain one-loop amplitude. We provide strong evidence that, in the bi-adjoint cubic scalar theory~\cite{Cachazo:2013iea}, forward limit on the two massive particles indeed leads to the correct one-loop amplitude in the massless sector. This is rather surprising at first sight: how can the massive propagators of tree amplitudes, like $1/(P^2-m^2)$, become any massless loop propagator $1/L^2$? 

The answer to this question is that they cannot! Instead, when taking the forward limit by setting the momenta of the two massive particles to $\pm\ell$ with $\ell^2=m^2$, the massive propagators in the tree become $1/((\ell-K)^2-\ell^2)$. As we re-interpret $\ell$ as the off-shell loop momentum, we will see that the tree amplitude exactly becomes the loop integrand in the new representation introduced in \cite{Geyer:2015bja}, which is given in terms of such factors. 
A crucial point of~\cite{Geyer:2015bja} is in realizing that this new form differs from the conventional form of one-loop integrals by terms that integrate to zero, so that it produces the correct one-loop amplitude after the loop integration. It is in this form that our forward limit works. This statement is independent of scattering equations or CHY formulas. Nevertheless, it leads to a closed formula based on the one-loop scattering equations, with an integrand that is completely parallel to its tree-level counterpart. 

The paper is organized as follows. In Section \ref{sec:formula}, by studying the forward limit of a scattering of massless particles with two massive particles, we derive the one-loop scattering equations, prove the counting of their solutions, present the general formula based on these equations and explain the condition for it to have a tree-level origin via forward limit. In Section  \ref{sec:amplitude} we argue that in the bi-adjoint scalar theory the forward limit of a tree amplitude with two massive particles is indeed a one-loop amplitude in the form of \cite{Geyer:2015bja}, which leads to a closed formula for the latter. We comment on several future directions in \ref{sec:discussion}.

{\it Note added}: While the manuscript is being prepared, the paper \cite{Baadsgaard:2015hia} appeared in arXiv, which proposed a different formula for the bi-adjoint scalar theory.

\section{One-loop Scattering Equations from the Forward Limit}\label{sec:formula}

In this section, we study the forward limit of Naculich's massive scattering equations~\cite{Naculich:2014naa,Naculich:2015zha}, which yields the $\SL2C$-covariant one-loop scattering equations. We provide a careful explanation of how the solutions and the measure behave in the limit. 

Let us first review Naculich's construction of scattering equations for $n$ massless particles with momenta $\{k_1,k_2,\ldots, k_n\}$, and two massive particles, labeled as ``$+$" and ``$-$" with momenta $k_+$ and $k_-$ respectively. The two massive ones share the same mass $m^2=k_+^2=k_-^2$, and there are $n+2$ massive scattering equations:
\ba\label{massivesca}
E_a:=&&\displaystyle \sum_{b=1, b\neq a}^n  \frac{k_a \cdot k_b}{\sigma_{a}-\sigma_{b}}+\frac{k_a\cdot k_+}{\sigma_a-\sigma_+}+\frac{k_a\cdot k_-}{\sigma_a-\sigma_-}=0\,, \quad a\in \{ 1,2,\ldots , n\};\nl
E_+:=&&\displaystyle~\sum_{b=1}^n \frac{k_+\cdot k_b}{\sigma_{+}-\sigma_{b}}+\frac{k_+ \cdot k_- +m^2}{\sigma_{+}-\sigma_{-}}=0\,,\quad E_-:=\displaystyle~\sum_{b=1}^n \frac{k_-\cdot k_b}{\sigma_{-}-\sigma_{b}}+\frac{k_+ \cdot k_- +m^2}{\sigma_{-}-\sigma_{+}}=0\,,
\ea
where $\sigma_{\pm}$ are the positions of the punctures for the two massive particles. The mass term in $k_+\cdot k_- +m^2$ is crucial for maintaining the $\SL2C$ symmetry~\cite{Cachazo:2013gna,Naculich:2014naa}, and again we fix positions of three punctures and remove three redundant equations. Exactly as the equations for $n{+}2$ massless particles, there are $(n-1)!$ solutions to the equations \eqref{massivesca}. The CHY formulas for tree amplitudes with $n$ massless particles and two massive particles read
\be\label{massiveformula}
{\cal M}^{\rm tree}_{n{+}2}(1,2,\ldots, n; +,-)=\int \frac{d\sigma_+ d\sigma_-d\,^n\sigma}{\textrm{vol}\,\SL2C} \prod_{a=1}^{n,+,-}{}'~\delta(E_a)~{\cal I}_{n{+}2}=:\int d\mu_{n{+}2}~{\cal I}_{n{+}2} \ee
where we used the abbreviation $d\mu_{n{+}2} $ for the measure including delta functions. 

The forward limit of the two massive particles amounts to taking $k^\mu_{\pm}\to \pm \ell^\mu$, where $\ell^2=m^2$. In the limit, the last term in $E_+$ and $E_-$ drops, since $k_+\cdot k_-=-\ell^2=-m^2$, and we arrive at the following equations which we refer to as ($\SL2C$-covariant) one-loop scattering equations
\ba\label{1loopsca}
&&{\cal E}_a:=\sum_{b=1, b\neq a}^n  \frac{k_a \cdot k_b}{\sigma_{a}-\sigma_{b}}+\frac{k_a\cdot \ell}{\sigma_a{-}\sigma_+}-\frac{k_a\cdot \ell}{\sigma_a{-}\sigma_-}=0\,,\quad a\in \{1,2, \ldots , n\};\nl
&&{\cal E}_+:=~\sum_{b=1}^n \frac{\ell \cdot k_b}{\sigma_+-\sigma_b}=0\,,\quad {\cal E}_-:=-\sum_{b=1}^n \frac{\ell \cdot k_b} {\sigma_--\sigma_b}=0\,.
\ea
By fixing $\sigma_+=0$, $\sigma_-=\infty$ and {\it e.g.} $\sigma_1=1$, and removing three redundant equations ${\cal E}_{\pm}$ and ${\cal E}_1$, the equations \eqref{1loopsca} reduce to the off-shell scattering equations \eqref{offsca}, introduced in~\cite{Geyer:2015bja}. 

While the equations \eqref{massivesca} seem to have a smooth limit, their $(n-1)!$ solutions do not, which are crucial in studying the behavior of CHY formulas in the limit. It turns out that the solutions fall into three sectors: a regular one, and two types of singular ones!

To illustrate this, let us stay in a neighborhood of the forward limit by assuming $k_++k_-=\tau q$, where $q$ is a fixed vector of finite norm and $\tau$ is a small parameter that controls the limit. Since $(k_+\cdot k_-+m^2)\sim\tau^2$, we expect the existence of solutions such that $(\sigma_5-\sigma_6)$ vanishes as $\tau\to0$, and we call them the singular solutions.

To analyze the singular solutions, it is convenient not to fix $\{\sigma_+,\sigma_-\}$, and to change variables $\sigma_\pm=z\pm\epsilon$. Since $\epsilon\to 0$ as $\tau\to 0$, we have $\epsilon\sim\tau^p$ for some $p>0$, and we can perturbatively solve the equations as expansions in $\tau$. At the leading order, note that in each $E_a$ the last two terms are irrelevant and thus they are equivalent to the tree-level massless equations~\eqref{scatt}, from which we obtain $(n-3)!$ solutions of $\{\sigma_a\}:=\{\sigma_1,\ldots,\sigma_n\}$. Given each solutions of $\{\sigma_a\}$, we then use the following two combinations to determine the leading order of $\{z, \epsilon\}$:
\begin{align}
\label{comb1}\epsilon(E_+-E_-)&=\sum_{b=1}^{n}\frac{(k_+-k_-)\cdot k_b}{z-\sigma_b}\epsilon+(k_+\cdot k_-+m^2)+\mathcal{O}(\epsilon^2)=0,\\
\label{comb2}E_++E_-&=\sum_{b=1}^{n}\frac{(k_++k_-)\cdot k_b}{z-\sigma_b}-\sum_{b=1}^{n}\frac{(k_+-k_-)\cdot k_b}{(z-\sigma_b)^2}\epsilon+\mathcal{O}(\epsilon^2)=0.
\end{align}

Now there are two different situations. Firstly,  if $p=1$, then at the leading order we have
\be
\sum_{b=1}^{n}\frac{(k_+-k_-)\cdot k_b}{z-\sigma_b}=0,\quad
\sum_{b=1}^{n}\frac{(k_++k_-)\cdot k_b}{z-\sigma_b}-\sum_{b=1}^{n}\frac{(k_+-k_-)\cdot k_b}{(z-\sigma_b)^2}\epsilon=0.
\ee
From the first equation it is obvious that given each $\{\sigma_a\}$ solution there are $(n-2)$ solutions for $\{z,\epsilon\}$. Secondly, if $p>1$, \eqref{comb1} forces $\epsilon\sim\tau^2$, and so at the leading order we have instead
\be
\sum_{b=1}^{n}\frac{(k_+-k_-)\cdot k_b}{z-\sigma_b}\epsilon+(k_+\cdot k_-+m^2)=0,\quad
\sum_{b=1}^{n}\frac{(k_++k_-)\cdot k_b}{z-\sigma_b}=0.
\ee
The second equation above leads to the same solution counting as the previous case.

Therefore, when approaching the forward limit, out of the $(n{-}1)!$ solutions of \eqref{massivesca}, there are two singular sectors in addition to the regular sector. This can be summarized as
\begin{center}
\begin{tabular}{@{}c|cc@{}}
\hline
sector & $\#$ of solutions & behavior \\
\hline
regular & ~$(n-1)!-2(n-2)!$~ & $|\sigma_+-\sigma_-|\sim1$ \\
singular I & $(n-2)!$ & $|\sigma_+-\sigma_-|\sim\tau$ \\
~singular II~ & $(n-2)!$ & ~$|\sigma_+-\sigma_-|\sim\tau^2$~ \\
\hline
\end{tabular}
\end{center}
In fact, the singular II sector drops out in the limit. Note that in this sector the term $\frac{k_+\cdot k_-+m^2}{\sigma_+-\sigma_-}\sim1$, hence it is relevant at the leading order of the equations $E_+$ and $E_-$. However, as we obtain $\mathcal{E}$'s in the strict limit \eqref{1loopsca}, this term is not present at all. We thus conclude that while there are $(n-1)!$ solutions to the massive scattering equations \eqref{massivesca}, the singular II sector is excluded in the forward limit; the solutions to the one-loop scattering equations \eqref{1loopsca} only come from the regular sector and the singular I sector, with a total number of $(n-1)!-(n-2)!$. We numerically checked this result up to six points; \textit{e.g.} for $n=4$, \eqref{1loopsca} has 2 regular solutions and 2 singular ones, and for $n=5$, it has $12$ regular solutions and 6 singular ones.  

In the forward limit, the tree-level formula \eqref{massiveformula} becomes a formula with one-loop scattering equations. However, the formula is only well-defined if \eqref{massiveformula} does not diverge and receives vanishing contribution from the singular II sector in the forward limit. One can easily see that the measure $d\mu_{n{+}2}$ in~\eqref{massiveformula} is of order $\tau^0$ for the singular I sector, while for the singular II sector it is of order $\tau^1$. Therefore, as long as the integrand ${\cal I}_{n{+}2}$ does not diverge as $\tau\to 0$, both conditions are satisfied and we have a well-defined formula in the limit. In that case $d\mu_{n{+}2}$ becomes the measure for the one-loop scattering equations:
\be\label{measure1loop}
d\mu_{n{+}2}\to d\mu^{(1)}_n:= \frac{d\sigma_+ d\sigma_-d\,^n\sigma}{\textrm{vol}\,\SL2C} \delta(\mathcal{E}_+)\delta(\mathcal{E}_-)\prod_a {}'\delta(\mathcal{E}_a)
\ee

There is a subtlety regarding the definition of the measure, due to the existence of the singular solutions (now we only have singular I). To avoid excluding their contribution, we are allowed to make any choice except for fixing $\{\sigma_+,\sigma_-\}$ at distinct locations. Moreover, as is explicitly presented in \eqref{measure1loop} we can only remove three arbitrary $\delta(\mathcal{E}_a)$ labeled by the particles but neither $\delta(\mathcal{E}_+)$ nor $\delta(\mathcal{E}_-)$. The reason is that although there are always three linear relations among the equations, when $\sigma_+=\sigma_-$ these relations are established among $\{\mathcal{E}_a\}$ only.

Recall that the version of one-loop scattering equations presented in \cite{Geyer:2015bja} (see \eqref{offsca}) are the gauge-fixed form of \eqref{1loopsca} by setting $\{\sigma_+,\sigma_-\}=\{0,\infty\}$ and $\sigma_1$ at some finite value; in the formula therein $\delta(\mathcal{E}_+)\delta(\mathcal{E}_-)\delta(\mathcal{E}_1)$ are removed, and as a consequence only the $(n-1)!-2(n-2)!$ regular solutions contribute. At first sight this seems to conflict the general formulation stated above. However, for certain integrand the singular I sector has vanishing contributions, then the two formulas of course give the same result. This is indeed the case for maximally super Yang--Mills, supergravity, and the $n$-gon formula studied in~\cite{Geyer:2015bja}.

\section{One-Loop Amplitude from Forward Limit: Bi-Adjoint Scalars}\label{sec:amplitude}

In the forward limit of a tree amplitude with two massive particles, the massive momenta become $k_\pm=\pm \ell$, which will be identified as the loop momentum.  Gluing the two legs together amounts to inserting a loop propagator $1/\ell^2$; one then has to sum over all internal states of the forward-limit particles, and perform the loop integral in the end. Our key question here is, does this procedure lead to the correct one-loop amplitudes? 
\be\label{question}
{\cal M}^{\rm 1-loop}_n(1,2,\ldots,n)~\stackrel{{?}}{=}~\int \frac{d^D \ell}{(2 \pi )^D} \sum_{{\rm int}_{\pm} }\frac {1} {\ell^2} \lim_{k_{\pm} \to \pm \ell}~{\cal M}_{n{+}2}^{\rm tree}(1,2,\ldots, n;+,-)\,,
\ee 
where on RHS we separate  $+,-$ from $1,2,\ldots,n$ to emphasize that they are massive.  

In this section, we argue that the answer to the above question is affirmative for the bi-adjoint cubic scalar theory, where the scalars are in the bi-adjoint of the group $U(N)\times U(\tilde N)$~\cite{Cachazo:2013iea}. To be more precise, the theory we consider has a universal cubic coupling for massless and massive scalars: for any tree amplitude with two massive scalars, it is given by a sum of Feynman diagrams with a unique massive line connecting them, while at one-loop level, we only consider massless amplitudes and the massive scalars do not appear in the Feynman diagrams. 

We will see that the whole procedure on the RHS of \eqref{question} produces exactly the sum of all one-loop Feynman integrals, including those with external-leg bubbles. The latter add up to the (massless) tree amplitude, multiplied by an integral which vanishes above the critical dimension of the theory,  $D=6$. The rest gives the amputated one-loop amplitude of the theory. As mentioned in the introduction, the precise result is in the representation of~\cite{Geyer:2015bja}, which differs from the conventional one at the integrand level but of course integrate to the correct one-loop amplitude.

\subsection{The Main Proposition for Bi-Adjoint Scalars}
\newcommand{\rad}{1.2cm}

For bi-adjoint scalars, the sum over internal states in~\eqref{question} amounts to the sum over (identified) color indices in $U(N)\times U(\tilde N)$, of the two massive scalars. This in turn implies that \eqref{question} can be recast in a simpler form, where both tree and one-loop amplitudes are decomposed into the so-called double-partial amplitudes introduced in~\cite{Cachazo:2013iea}. It is convenient to use the color-decomposition of~\cite{DelDuca:1999rs}: for ${\cal M}^{\rm tree}_{n{+}2}$, one can choose the color basis of $U(N)$ (or $U(\tilde N)$) to be the $n!$ half-ladder diagrams, with $+$ and $-$ fixed on the two ends. With the sum over their colors in the forward limit, the two ends are glued and we recover exactly the one-loop color basis for ${\cal M}_n^{\rm 1-loop}$, which are $(n-1)!$ ring diagrams~\cite{DelDuca:1999rs}.

Then it is easy to show that the validity of \eqref{question} is completely equivalent to the validity of the following relations involving double-partial amplitudes:
\be\label{partialamp}
\boxed{m^{\rm 1-loop}_n[\pi|\rho]~{=}~\int \frac{d^D \ell}{(2 \pi )^D}~\frac {1} {\ell^2} \lim_{k_{\pm} \to \pm \ell}~\sum_{\substack{\alpha\in {\rm cyc}(\pi)\\ \beta\in {\rm cyc}(\rho)}}\hspace{-.5em}m_{n{+}2}^{\rm tree}[-\alpha+|-\beta+]\,.}
\ee
where $\pi,\rho \in S_n/Z_n$ are permutations of ring diagrams, defined modulo cyclic shifts, and $m^{\rm 1-loop}_n[\pi | \rho]$ are one-loop double-partial amplitudes; $m_{n{+}2}^{\rm tree}[-\alpha+|-\beta+]$ are tree double-partial amplitudes with the positions of massive particles, $+$ and $-$ fixed\footnote{Here we choose the convention that the loop momentum flows in the same direction as the ordering $\pi$ ($\rho$) goes when we view the Feynman diagrams in $m_n^{1\text{-loop}}[\pi|\rho]$ as embedded in a plane. If it is in the opposite direction then on the RHS of \eqref{partialamp} we have $m_{n+2}^{\rm tree}[+\alpha-|+\beta-]$. The other two possibilities of inserting the $(+-)$ pair are related to these, \textit{e.g.} $m_{n+2}^{\rm tree}[-\alpha+|+\beta-]=(-1)^nm_{n+2}^{\rm tree}[-\alpha+|-\beta^{-1}+]$, where $\beta^{-1}$ denotes the reverse ordering of $\beta$.}. The sum over the permutation $\alpha\in S_n$ ($\beta\in S_n$) is over the $n$ cyclic images of $\pi$ ($\rho$), which correspond to exactly the $n$ half-ladder diagrams contributing to the same ring diagram. 

Eq.~\eqref{partialamp} is our main proposition on the forward limit of massive trees, and its validity leads to a formula with one-loop scattering equations. Recall the CHY formula for tree-level double-partial amplitudes, $m^{\rm tree}_{n{+}2}[-\alpha+|-\beta+]$~\cite{Cachazo:2013iea}:
\be
m^{\rm tree}_{n{+}2}[-\alpha+|-\beta+]=\int d\mu_{n{+}2}~PT(-, \alpha(1), \ldots, \alpha(n), +)~PT(-, \beta(1), \ldots, \beta(n), +)\,,
\ee
where the Parke-Taylor factor for a permutation of labels $\{1,2,\ldots,n\}$ is defined as
\be
PT(-,\alpha(1),\ldots,\alpha(n),+)=\frac{1}{(\sigma_--\sigma_{\alpha(1)}) (\sigma_{\alpha(1)}-\sigma_{\alpha(2)}) \cdots (\sigma_{\alpha(n)}-\sigma_+)(\sigma_+-\sigma_-)}\,.
\ee
Plug this to the RHS of \eqref{partialamp}, then by~\eqref{measure1loop} we obtain a formula for $m^{\rm 1-loop}_n[\pi|\rho]$, 
\be\label{1loopphi3}
m^{\rm 1-loop}_n [\pi|\rho]=\int \frac{d^D \ell}{(2 \pi )^D}~\frac {1} {\ell^2} \int d\mu^{(1)}_n \sum_{\alpha\in {\rm cyc}(\pi)}PT(-,\alpha,+)~\sum_{\beta\in {\rm cyc}(\rho)} PT(-, \beta,+)\,.
\ee
In order that \eqref{1loopphi3} is well-defined, one need to ensure that the integrand in \eqref{1loopphi3} does not diverge on the singular solutions. This is guaranteed here because the residue of, \textit{e.g.} $\sum_\alpha PT(-,\alpha,+)$ on $\sigma_+=\sigma_-$ vanishes by the $\text{U}(1)$ decoupling identity, and so this integrand remains finite. 

An immediate consequence of \eqref{1loopphi3} is a formula for a $n$-gon integral of any given ordering, $\pi$. From the color-decomposition of \cite{DelDuca:1999rs} mentioned above, it is given by the sum of double-partial amplitudes over all inequivalent orderings $\rho$ with $\pi$ fixed, and \eqref{partialamp} yields
\be\label{ngon}\begin{split}
n\text{-gons}(\pi) &= \sum_{\rho\in S_{n{-}1}} m^{\rm 1-loop}[\pi|\rho]\\
&=\int \frac{d^D \ell}{(2 \pi )^D}~\frac {1} {\ell^2} \int d\mu^{(1)}_n \sum_{\alpha\in {\rm cyc}(\pi)}PT(-,\alpha,+)~{\rm sym} (1,2,\ldots,n; +,-)\,,
\end{split}
\ee
where in the second equality we have plugged \eqref{1loopphi3} in, and defined the sum of Parke-Taylor factors which is totally symmetric in labels $1,2,\ldots,n$:
\be
 {\rm sym} (1,2,\ldots,n; +,-)=\sum_{\rho\in S_{n{-}1}} ~\sum_{\beta\in {\rm cyc}(\rho)} PT(-, \beta,+)
=\sum_{\beta\in S_n} PT(-, \beta,+)\,.
\ee
Furthermore, following~\cite{Geyer:2015bja}, one can define the sum of $(n-1)!$ inequivalent $n$-gon integrals (let us denote it as ``sym. $n$-gons"), which is given by a remarkably simple formula:
\be\label{ngons}
{\rm sym.}~n\text{-gons}:=\sum_{\pi\in S_{n{-}1}} n\text{-gons} (\pi)=\int \frac{d^D \ell}{(2 \pi )^D}~\frac {1} {\ell^2} \int d\mu^{(1)}_n {\rm sym} (1,2,\ldots,n; +,-)^2\,.
\ee

An important feature of both \eqref{ngon} and \eqref{ngons} is that all singular I solutions have vanishing contributions, thus only the $(n-1)!-2(n-2)!$ regular solutions contribute to the formula. As mentioned before, in such cases one can do the gauge fixing that gives \eqref{offsca}: $\sigma_+=0$, $\sigma_-=\infty$, and it is straightforward to obtain an even simpler form of sym$(1,2,\ldots,n;+,-)$:
\be
 {\rm sym} (1,2,\ldots,n; +,-)|_{\sigma_+=0, \sigma_-=\infty}\sim \prod_{i=1}^n \frac{1}{\sigma_i},
\ee 
where $\sim$ means that we have compensated by the factor $(\sigma_-)^2$, from gauge fixing, and this reproduces the so-called $n$-gon formula of~\cite{Geyer:2015bja}. This belongs to the special cases where only regular solutions contribute; but for general cases, such as \eqref{partialamp}, one has to include all the $(n-1)!-(n-2)!$ solutions to the one-loop scattering equations. 

While \eqref{partialamp} leads to a CHY formula for one-loop bi-adjoint scalar amplitudes, \eqref{1loopphi3}, the proposition itself is simply a statement about one-loop and tree amplitudes. Let us first spell out the LHS and especially the new representation introduced in~\cite{Geyer:2015bja}. Similar to the discussion at tree level in~\cite{Cachazo:2013iea}, we define the set of one-loop trivalent scalar diagrams respecting the planar ordering $\pi$ as $T^{(1)}(\pi)$. Then the one-loop double-partial amplitudes are given by the sum of all diagrams respecting both $\pi$ and $\rho$ orderings:
\be\label{1loopFD}
m^{\rm 1-loop}_n [\pi|\rho]=(-)^{n_{\pi|\rho}}~\int \frac{d^D L}{(2 \pi )^D} \sum_{g\in T^{(1)}(\pi)\cap T^{(1)}(\rho)}~\prod_{e\in E(g)} \frac {1}{P_e(L)}\,,
\ee
where the overall sign is determined by the number of flips between two permutations $\pi,\rho$, and $1/P_e(L)$ are propagators of a diagram $g$ ($L$ is the loop momentum). In general we encounter diagrams with tadpoles, bubbles, triangles, ..., up to $n$-gons, with possible tree structures attached to the corners. Let us recast these scalar integrals in the new representation in \cite{Geyer:2015bja}, which does not affect the tree structures. For any scalar diagram, after chopping off the tree structures, the left-over piece of the loop integrand is an $m$-gon scalar integrand of the form
\be
\tilde{I}_m(L; K_1,\ldots, K_m):=\frac {1}{L^2 (L-K_1)^2 (L-K_1-K_2)^2~\cdots~(L-K_1-\ldots-K_{m{-}1})^2}\,,
\ee
where $m=2,\ldots,n$, with the total momentum at its $i^{\rm th}$ corner denoted as $K_i$ (thus possibly massive). By applying the partial fraction identity $1/(A B)=1/(A (A{-}B))+1/ (B (B{-}A))$, one can rewrite $\tilde{I}_m$ uniquely into $m$ terms, such that each term has one of the original propagators and $m{-}1$ factors given by differences. Note that for each individual term, we can shift the loop momentum without changing the result of the integral; thus we redefine the only propagator to be $\ell^2$, and the rest factors take the form $(\ell-\sum K)^2-\ell^2$. In the end we obtain another integrand
\be\label{newrep}
I_m(\ell; K_1,\ldots, K_m):=\displaystyle \frac{1}{\ell^2}~\sum_{a=1}^m~\prod_{b=a}^{a{+}m{-}2} \frac{1}{(\ell-K_a-K_{a{+}1}-\ldots-K_b)^2-\ell^2}\,,
\ee
where all labels are understood as mod $m$, and except for $\ell^2$ the other $m{-}1$ factors in a denominator are linear in $\ell$. The two integrands obviously integrate to the same result
\be\label{looptreeid}
\int \frac{d^D L}{(2 \pi )^D} \tilde{I}_m(L; K_1,\ldots, K_m)=\int \frac{d^D \ell}{(2 \pi )^D} I_m(\ell; K_1,\ldots, K_m)\,.
\ee
As some examples, the new representation for bubbles, triangles, and boxes are given by
\ba
I_2&=&\frac{1}{\ell^2 ((\ell-K_1)^2-\ell^2)}+\frac{1}{\ell^2((\ell-K_2)^2-\ell^2)}\,\\
I_3&=&\frac{1}{\ell^2 ((\ell-K_1)^2-\ell^2)~((\ell+K_3)^2-\ell^2)}+{\rm cyc}(K_1,K_2,K_3)\,,\nl
I_4&=&\frac{1}{\ell^2 ((\ell-K_1)^2-\ell^2)~((\ell-K_1-K_2)^2-\ell^2)~((\ell+K_4)^2-\ell^2)}+{\rm cyc}(K_1,K_2,K_3,K_4)\,.\nonumber
\ea

Note that each term in the summation on the RHS of \eqref{newrep} can be interpreted as a single ladder-like tree diagram, with a massive line running in the ladder, whose two ends have the on-shell momenta $+\ell$ and $-\ell$ due to the momentum conservation $K_1+K_2+\cdots+K_m=0$. Hence diagrammatically the identity \eqref{looptreeid} means
\begin{equation}\label{looptreegraph}
\parbox{1.8cm}{\tikz{
\begin{scope}[scale=.7]
\node [rotate=-60] at (210:\rad) {$\cdots$};
\draw [very thick, black] (30:5*\rad/7) -- (30:\rad);
\draw [very thick, black] (90:5*\rad/7) -- (90:\rad);
\draw [very thick, black] (150:5*\rad/7) -- (150:\rad);
\draw [very thick, black] (270:5*\rad/7) -- (270:\rad);
\draw [very thick, black] (330:5*\rad/7) -- (330:\rad);
\draw [very thick, black,->] (0:5*\rad/7) arc [start angle=0, end angle=370, radius=5*\rad/7];
\end{scope}
}}=\int\frac{d^D\ell}{(2\pi)^D}\frac{1}{\ell^2}\lim_{{\rm\color{SeaGreen}forward}}\left(\!
\parbox{1.8cm}{\tikz{
\begin{scope}[scale=.7]
\node [rotate=-60] at (210:\rad) {$\cdots$};
\draw [very thick, black] (30:5*\rad/7) -- (30:\rad);
\draw [very thick, black] (90:5*\rad/7) -- (90:\rad);
\draw [very thick, black] (150:5*\rad/7) -- (150:\rad);
\draw [very thick, black] (270:5*\rad/7) -- (270:\rad);
\draw [very thick, black] (330:5*\rad/7) -- (330:\rad);
\draw [very thick, SeaGreen] (10:5*\rad/7) arc [start angle=10, end angle=350, radius=5*\rad/7];
\draw [very thick, black] (30:5*\rad/7) arc [start angle=30, end angle=330, radius=5*\rad/7];
\end{scope}
}}+\!
\parbox{1.8cm}{\tikz{
\begin{scope}[scale=.7]
\node [rotate=-60] at (210:\rad) {$\cdots$};
\draw [very thick, black] (30:5*\rad/7) -- (30:\rad);
\draw [very thick, black] (90:5*\rad/7) -- (90:\rad);
\draw [very thick, black] (150:5*\rad/7) -- (150:\rad);
\draw [very thick, black] (270:5*\rad/7) -- (270:\rad);
\draw [very thick, black] (330:5*\rad/7) -- (330:\rad);
\draw [very thick, SeaGreen] (70:5*\rad/7) arc [start angle=70, end angle=410, radius=5*\rad/7];
\draw [very thick, black] (90:5*\rad/7) arc [start angle=90, end angle=390, radius=5*\rad/7];
\end{scope}
}}+\!
\parbox{1.8cm}{\tikz{
\begin{scope}[scale=.7]
\node [rotate=-60] at (210:\rad) {$\cdots$};
\draw [very thick, black] (30:5*\rad/7) -- (30:\rad);
\draw [very thick, black] (90:5*\rad/7) -- (90:\rad);
\draw [very thick, black] (150:5*\rad/7) -- (150:\rad);
\draw [very thick, black] (270:5*\rad/7) -- (270:\rad);
\draw [very thick, black] (330:5*\rad/7) -- (330:\rad);
\draw [very thick, SeaGreen] (130:5*\rad/7) arc [start angle=130, end angle=470, radius=5*\rad/7];
\draw [very thick, black] (150:5*\rad/7) arc [start angle=150, end angle=450, radius=5*\rad/7];
\end{scope}
}}+\cdots+\!
\parbox{1.8cm}{\tikz{
\begin{scope}[scale=.7]
\node [rotate=-60] at (210:\rad) {$\cdots$};
\draw [very thick, black] (30:5*\rad/7) -- (30:\rad);
\draw [very thick, black] (90:5*\rad/7) -- (90:\rad);
\draw [very thick, black] (150:5*\rad/7) -- (150:\rad);
\draw [very thick, black] (270:5*\rad/7) -- (270:\rad);
\draw [very thick, black] (330:5*\rad/7) -- (330:\rad);
\draw [very thick, SeaGreen] (-70:5*\rad/7) arc [start angle=-70, end angle=-410, radius=5*\rad/7];
\draw [very thick, black] (-90:5*\rad/7) arc [start angle=-90, end angle=-390, radius=5*\rad/7];
\end{scope}
}}\right),
\end{equation}
where we take the forward limit on the two massive particles at the ends of each ladder. This fact essentially allows us to show \eqref{partialamp}: by recasting \eqref{1loopFD} in the new representation \eqref{newrep}, we expect the result to be identical to the RHS of \eqref{partialamp}. As we will see shortly, the proof can be formulated as a diagrammatic rule, which assembles tree-level cubic diagrams into the RHS of \eqref{looptreegraph}. For general $n$, the complete proof with all the combinatorics worked out is involved and not very illuminating. However, the $n=4$ results already illustrate all the features of the proof, and we will present them in full details. 

Two comments here are in order. First, already from the antisymmetry of the structure constant of cubic vertex, all tadpole diagrams must cancel in the sum over diagrams, so effectively no tadpoles appear in the one-loop result. We will see shortly that the same is true from the RHS of \eqref{partialamp}. In addition, it is straightforward to show that all diagrams with external-leg bubble add up to the $n$-point tree amplitude, multiplied by an overall integral:
\be
\Delta_n~{\cal M}^{\rm tree}_n:=-\frac{1}{4} \int \frac{d^D \ell}{(2 \pi )^D} \frac{1}{\ell^2}\sum_{a=1}^{n}\frac{1}{(\ell\cdot k_a)^2} \times {\cal M}^{\rm tree}_n\,.
\ee
This universal term is related to the renormalization of one-loop amplitudes, and we will see that exactly the same term appears on RHS of \eqref{partialamp}. In dimensions $D>6$, the integral $\Delta_n$ vanishes, and the rest is the sum of all amputated Feynman diagrams at one loop.

\subsection{Four-Point Results: Diagrammatic Rules from Tree to One-Loop}

\newcommand{\dotsize}{1.5pt}
\tikzset{
edot/.style={circle,draw=black,thick,fill=black,inner sep=\dotsize},
idot/.style={circle,draw=black,thick,fill=white,inner sep=\dotsize}
}

Here we provide a detailed proof of \eqref{partialamp} for $n=4$, which illustrate the diagrammatic rule needed for taking the forward limit of tree amplitudes on the RHS of \eqref{partialamp} and producing one-loop amplitudes on LHS. We will not present the complete proof of the proposition, but it should follow similarly from our discussions on 
these examples.

In principle there are $3!\times 3!=36$ different $m_4^{\rm 1-loop} [\pi|\rho]$, but most of them are related to each other by relabeling, and there are only three independent ones that we have to study. Obviously we can fix $\pi=(1234)$, and we choose
\begin{equation}
[\pi|\rho]\in\{[1234|1243],[1234|1234],[1234|1432]\},
\end{equation}
since the other three choices $\rho\in\{(1324),(1342),(1423)\}$ are related to $\rho=(1243)$ by cyclic permutations of the four labels. 

Let us show \eqref{partialamp} for the first example now:
\begin{equation}\label{example1}
m_4^{1\text{-loop}}[1234|1243]=\int\frac{d^D\ell}{(2\pi)^D}\frac{1}{\ell^2}\lim_{k_\pm\to\pm\ell}
\sum_{\substack{\alpha\in\text{cyc}(1234)\\\beta\in\text{cyc}(1243)}}\hspace{-.5em}m_6^{\rm tree}[-\alpha+|-\beta+]\,,
\end{equation}
A generic $m_n^{\rm tree}$ is reproduced by the corresponding CHY formula in \cite{Cachazo:2013iea}, which comes with a simple graphic interpretation of $m_n^{\rm tree}$, from which one can directly read off the result, including the overall sign. This tool turns out to be useful for our purpose here.

The sum on RHS is equivalent to inserting the ordered pair $(+-)$ into the original cyclic ordering $(1234)$ and $(1243)$ respectively. We will first consider the sum over $\beta$ (for any given $\alpha$) and then sum over $\alpha$. The object before inserting $(+-)$ is the four-point double-partial amplitude $m_4^{\rm tree}[\pi|\rho]\equiv m_4^{\rm tree}[1234|1243]$. Recall the graphic representation introduced in~\cite{Cachazo:2013iea}, we have

\begin{equation}\label{4pttree}
m_4^{\rm tree}[{\color{OrangeRed}1234}|{\color{Cyan}1243}]=
\parbox{3cm}{\tikz{
\draw [thick,OrangeRed] (0,0) circle [radius=\rad];
\coordinate [label=-30:$1$] (p1) at (-30:\rad);
\coordinate [label=+30:$2$] (p2) at (+30:\rad);
\coordinate [label=+150:$3$] (p3) at (+150:\rad);
\coordinate [label=-150:$4$] (p4) at (-150:\rad);
\draw [very thick,Cyan] (p1) -- (p2) -- (p4) -- (p3) -- cycle;
\node [edot] at (p1) {};
\node [edot] at (p2) {};
\node [edot] at (p3) {};
\node [edot] at (p4) {};
}}=
\parbox{2.5cm}{\tikz{
\begin{scope}[scale=.8]
\draw [very thick,black] (-30:\rad) -- (0:\rad/3) -- (30:\rad);
\draw [very thick,black] (-150:\rad) -- (180:\rad/3) -- (150:\rad);
\draw [very thick,black] (0:\rad/3) -- (180:\rad/3);
\coordinate [label=-30:$1$] (p1) at (-30:\rad);
\coordinate [label=+30:$2$] (p2) at (+30:\rad);
\coordinate [label=+150:$3$] (p3) at (+150:\rad);
\coordinate [label=-150:$4$] (p4) at (-150:\rad);
\end{scope}
}}=\frac{1}{(k_1+k_2)^2}\,.
\end{equation}
The correspondence between the graph and the tree diagram above can be roughly quoted as that each blue triangle indicates a trivalent vertex, and that the two trivalent vertices are glued together in the same way as the triangles.

In the above graph the four dots divide the red disk boundary into four sides; 
for each given $\alpha\in\text{cyc}(\pi)$, the sum $\sum_\beta m_6^{\rm tree}[-\alpha+|-\beta+]$ can be represented as inserting the ordered pair of points $(+ -)$ on one of the sides. For each term in this summation, we replace one of the blue segments by connecting its two end points to the points $+$ and $-$ according to the ordering in $(\beta + -)$. The resulting graph is a graph representation of the corresponding $m_6^{\rm tree}$, which immediately tells what diagrams are contained. For example,
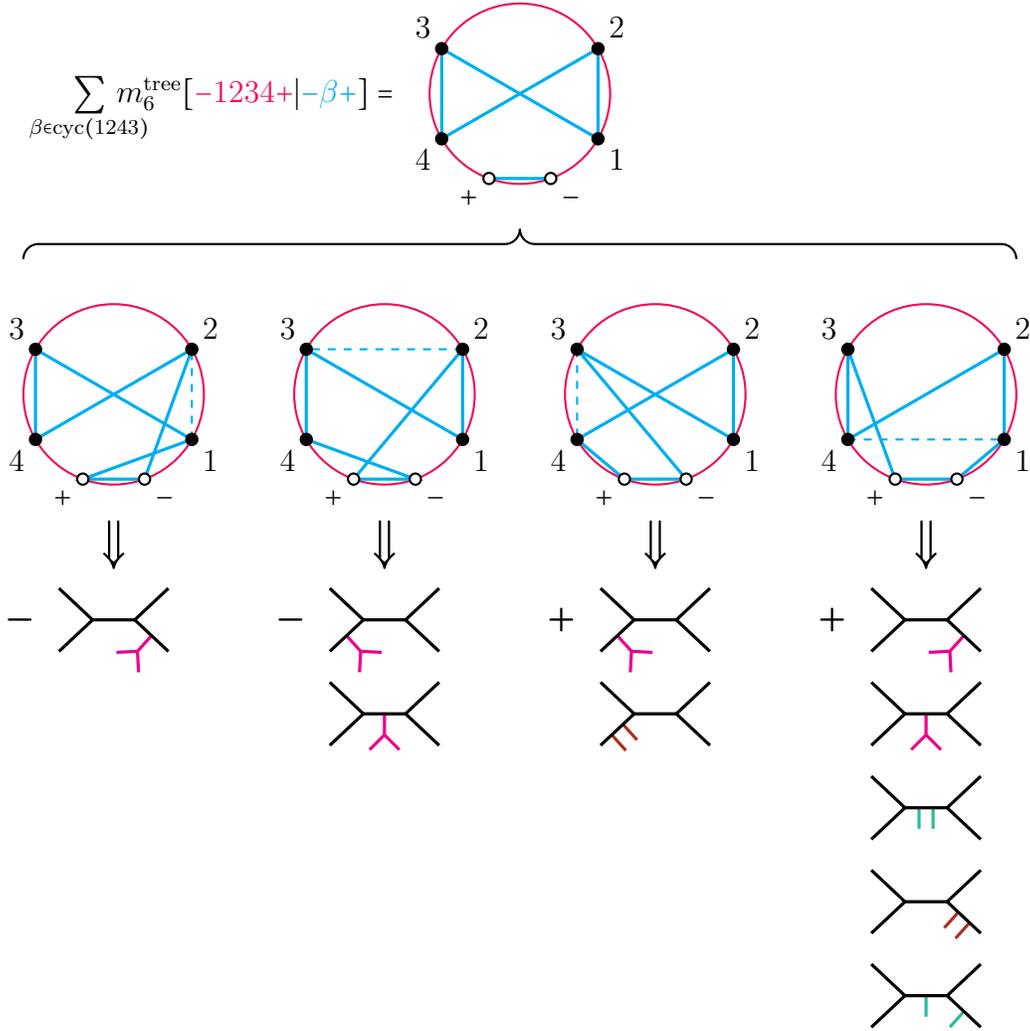
\begin{center}
\begin{tikzpicture}
\node [xshift=-4.1cm,yshift=-.2cm] at (0,0) {$\displaystyle\sum_{\beta\in\text{cyc}(1243)}\hspace{-.5cm}m_6^{\rm tree}[{\color{OrangeRed}-1234+}|{\color{Cyan}-\beta+}]=$};
\begin{scope}
\draw [thick,OrangeRed] (0,0) circle [radius=\rad];
\coordinate [label=-30:$1$] (p1) at (-30:\rad);
\coordinate [label=+30:$2$] (p2) at (+30:\rad);
\coordinate [label=+150:$3$] (p3) at (+150:\rad);
\coordinate [label=-150:$4$] (p4) at (-150:\rad);
\coordinate [label=-70:$-$] (minus) at (-70:\rad);
\coordinate [label=-110:$+$] (plus) at (-110:\rad);
\draw [very thick,Cyan] (p1) -- (p2) -- (p4) -- (p3) -- cycle;
\draw [very thick,Cyan] (plus) -- (minus);
\node [edot] at (p1) {};
\node [edot] at (p2) {};
\node [edot] at (p3) {};
\node [edot] at (p4) {};
\node [idot] at (minus) {};
\node [idot] at (plus) {};
\end{scope}
\begin{scope}[thick,black]
\draw (0,-1.8cm) arc [start angle=0, end angle=-90, radius=.2cm] -- (-6.4cm,-2cm) arc [start angle=90, end angle=180, radius=.2cm];
\draw (0,-1.8cm) arc [start angle=180, end angle=270, radius=.2cm] -- (6.4cm,-2cm) arc [start angle=90, end angle=0, radius=.2cm];
\end{scope}
\begin{scope}[xshift=-5.4cm,yshift=-4cm]
\draw [thick,OrangeRed] (0,0) circle [radius=\rad];
\coordinate [label=-30:$1$] (p1a) at (-30:\rad);
\coordinate [label=+30:$2$] (p2a) at (+30:\rad);
\coordinate [label=+150:$3$] (p3a) at (+150:\rad);
\coordinate [label=-150:$4$] (p4a) at (-150:\rad);
\coordinate [label=-70:$-$] (minusa) at (-70:\rad);
\coordinate [label=-110:$+$] (plusa) at (-110:\rad);
\draw [very thick,Cyan] (p1a) -- (plusa) -- (minusa) -- (p2a) -- (p4a) -- (p3a) -- cycle;
\draw [thick,dashed,Cyan] (p1a) -- (p2a);
\node [edot] at (p1a) {};
\node [edot] at (p2a) {};
\node [edot] at (p3a) {};
\node [edot] at (p4a) {};
\node [idot] at (minusa) {};
\node [idot] at (plusa) {};
\end{scope}
\begin{scope}[xshift=-1.8cm,yshift=-4cm]
\draw [thick,OrangeRed] (0,0) circle [radius=\rad];
\coordinate [label=-30:$1$] (p1b) at (-30:\rad);
\coordinate [label=+30:$2$] (p2b) at (+30:\rad);
\coordinate [label=+150:$3$] (p3b) at (+150:\rad);
\coordinate [label=-150:$4$] (p4b) at (-150:\rad);
\coordinate [label=-70:$-$] (minusb) at (-70:\rad);
\coordinate [label=-110:$+$] (plusb) at (-110:\rad);
\draw [very thick,Cyan] (p1b) -- (p2b) -- (plusb) -- (minusb) -- (p4b) -- (p3b) -- cycle;
\draw [thick,dashed,Cyan] (p2b) -- (p3b);
\node [edot] at (p1b) {};
\node [edot] at (p2b) {};
\node [edot] at (p3b) {};
\node [edot] at (p4b) {};
\node [idot] at (minusb) {};
\node [idot] at (plusb) {};
\end{scope}
\begin{scope}[xshift=+1.8cm,yshift=-4cm]
\draw [thick,OrangeRed] (0,0) circle [radius=\rad];
\coordinate [label=-30:$1$] (p1c) at (-30:\rad);
\coordinate [label=+30:$2$] (p2c) at (+30:\rad);
\coordinate [label=+150:$3$] (p3c) at (+150:\rad);
\coordinate [label=-150:$4$] (p4c) at (-150:\rad);
\coordinate [label=-70:$-$] (minusc) at (-70:\rad);
\coordinate [label=-110:$+$] (plusc) at (-110:\rad);
\draw [very thick,Cyan] (p1c) -- (p2c) -- (p4c) -- (plusc) -- (minusc) -- (p3c) -- cycle;
\draw [thick,dashed,Cyan] (p3c) -- (p4c);
\node [edot] at (p1c) {};
\node [edot] at (p2c) {};
\node [edot] at (p3c) {};
\node [edot] at (p4c) {};
\node [idot] at (minusc) {};
\node [idot] at (plusc) {};
\end{scope}
\begin{scope}[xshift=+5.4cm,yshift=-4cm]
\draw [thick,OrangeRed] (0,0) circle [radius=\rad];
\coordinate [label=-30:$1$] (p1d) at (-30:\rad);
\coordinate [label=+30:$2$] (p2d) at (+30:\rad);
\coordinate [label=+150:$3$] (p3d) at (+150:\rad);
\coordinate [label=-150:$4$] (p4d) at (-150:\rad);
\coordinate [label=-70:$-$] (minusd) at (-70:\rad);
\coordinate [label=-110:$+$] (plusd) at (-110:\rad);
\draw [very thick,Cyan] (p1d) -- (p2d) -- (p4d) -- (p3d) -- (plusd) -- (minusd) -- cycle;
\draw [thick,dashed,Cyan] (p4d) -- (p1d);
\node [edot] at (p1d) {};
\node [edot] at (p2d) {};
\node [edot] at (p3d) {};
\node [edot] at (p4d) {};
\node [idot] at (minusd) {};
\node [idot] at (plusd) {};
\end{scope}
\begin{scope}[yshift=-6cm]
\node [scale=1.6] at (-5.4cm,0) {$\Downarrow$};
\node [scale=1.6] at (-1.8cm,0) {$\Downarrow$};
\node [scale=1.6] at (+1.8cm,0) {$\Downarrow$};
\node [scale=1.6] at (+5.4cm,0) {$\Downarrow$};
\end{scope}
\begin{scope}[xshift=-1.25cm,yshift=-7cm]
\node [scale=1.6] at (-5.4cm,0) {$-$};
\node [scale=1.6] at (-1.8cm,0) {$-$};
\node [scale=1.6] at (+1.8cm,0) {$+$};
\node [scale=1.6] at (+5.4cm,0) {$+$};
\end{scope}
\begin{scope}[xshift=-5.4cm,yshift=-7cm,scale=.7]
\coordinate (p1) at (-30:\rad);
\coordinate (p2) at (+30:\rad);
\coordinate (p3) at (+150:\rad);
\coordinate (p4) at (-150:\rad);
\coordinate (i1) at (0:\rad/3);
\coordinate (i2) at (180:\rad/3);
\draw [very thick,Magenta] ($(i1)!.5!(p1)$) -- ++(-132:\rad/3) -- ++(-132-45:\rad/3);
\draw [very thick,Magenta] ($(i1)!.5!(p1)$) ++(-132:\rad/3) -- ++(-132+45:\rad/3);
\draw [very thick,black] (p1) -- (i1) -- (p2);
\draw [very thick,black] (p3) -- (i2) -- (p4);
\draw [very thick,black] (i1) -- (i2);
\end{scope}
\begin{scope}[xshift=-1.8cm,yshift=-7cm,scale=.7]
\coordinate (p1) at (-30:\rad);
\coordinate (p2) at (+30:\rad);
\coordinate (p3) at (+150:\rad);
\coordinate (p4) at (-150:\rad);
\coordinate (i1) at (0:\rad/3);
\coordinate (i2) at (180:\rad/3);
\draw [very thick,Magenta] ($(i2)!.5!(p4)$) -- ++(-48:\rad/3) -- ++(-48-45:\rad/3);
\draw [very thick,Magenta] ($(i2)!.5!(p4)$) ++(-48:\rad/3) -- ++(-48+45:\rad/3);
\draw [very thick,black] (p1) -- (i1) -- (p2);
\draw [very thick,black] (p3) -- (i2) -- (p4);
\draw [very thick,black] (i1) -- (i2);
\end{scope}
\begin{scope}[xshift=-1.8cm,yshift=-8.25cm,scale=.7]
\coordinate (p1) at (-30:\rad);
\coordinate (p2) at (+30:\rad);
\coordinate (p3) at (+150:\rad);
\coordinate (p4) at (-150:\rad);
\coordinate (i1) at (0:\rad/3);
\coordinate (i2) at (180:\rad/3);
\draw [very thick,Magenta] ($(i1)!.5!(i2)$) -- ++(-90:\rad/3) -- ++(-90-45:\rad/3);
\draw [very thick,Magenta] ($(i1)!.5!(i2)$) ++(-90:\rad/3) -- ++(-90+45:\rad/3);
\draw [very thick,black] (p1) -- (i1) -- (p2);
\draw [very thick,black] (p3) -- (i2) -- (p4);
\draw [very thick,black] (i1) -- (i2);
\end{scope}
\begin{scope}[xshift=1.8cm,yshift=-7cm,scale=.7]
\coordinate (p1) at (-30:\rad);
\coordinate (p2) at (+30:\rad);
\coordinate (p3) at (+150:\rad);
\coordinate (p4) at (-150:\rad);
\coordinate (i1) at (0:\rad/3);
\coordinate (i2) at (180:\rad/3);
\draw [very thick,Magenta] ($(i2)!.5!(p4)$) -- ++(-48:\rad/3) -- ++(-48-45:\rad/3);
\draw [very thick,Magenta] ($(i2)!.5!(p4)$) ++(-48:\rad/3) -- ++(-48+45:\rad/3);
\draw [very thick,black] (p1) -- (i1) -- (p2);
\draw [very thick,black] (p3) -- (i2) -- (p4);
\draw [very thick,black] (i1) -- (i2);
\end{scope}
\begin{scope}[xshift=1.8cm,yshift=-8.25cm,scale=.7]
\coordinate (p1) at (-30:\rad);
\coordinate (p2) at (+30:\rad);
\coordinate (p3) at (+150:\rad);
\coordinate (p4) at (-150:\rad);
\coordinate (i1) at (0:\rad/3);
\coordinate (i2) at (180:\rad/3);
\draw [very thick,Mahogany] ($(i2)!1/3!(p4)$) -- ++(-48:\rad/3);
\draw [very thick,Mahogany] ($(i2)!2/3!(p4)$) -- ++(-48:\rad/3);
\draw [very thick,black] (p1) -- (i1) -- (p2);
\draw [very thick,black] (p3) -- (i2) -- (p4);
\draw [very thick,black] (i1) -- (i2);
\end{scope}
\begin{scope}[xshift=5.4cm,yshift=-7cm,scale=.7]
\coordinate (p1) at (-30:\rad);
\coordinate (p2) at (+30:\rad);
\coordinate (p3) at (+150:\rad);
\coordinate (p4) at (-150:\rad);
\coordinate (i1) at (0:\rad/3);
\coordinate (i2) at (180:\rad/3);
\draw [very thick,Magenta] ($(i1)!.5!(p1)$) -- ++(-132:\rad/3) -- ++(-132-45:\rad/3);
\draw [very thick,Magenta] ($(i1)!.5!(p1)$) ++(-132:\rad/3) -- ++(-132+45:\rad/3);
\draw [very thick,black] (p1) -- (i1) -- (p2);
\draw [very thick,black] (p3) -- (i2) -- (p4);
\draw [very thick,black] (i1) -- (i2);
\end{scope}
\begin{scope}[xshift=5.4cm,yshift=-8.25cm,scale=.7]
\coordinate (p1) at (-30:\rad);
\coordinate (p2) at (+30:\rad);
\coordinate (p3) at (+150:\rad);
\coordinate (p4) at (-150:\rad);
\coordinate (i1) at (0:\rad/3);
\coordinate (i2) at (180:\rad/3);
\draw [very thick,Magenta] ($(i1)!.5!(i2)$) -- ++(-90:\rad/3) -- ++(-90-45:\rad/3);
\draw [very thick,Magenta] ($(i1)!.5!(i2)$) ++(-90:\rad/3) -- ++(-90+45:\rad/3);
\draw [very thick,black] (p1) -- (i1) -- (p2);
\draw [very thick,black] (p3) -- (i2) -- (p4);
\draw [very thick,black] (i1) -- (i2);
\end{scope}
\begin{scope}[xshift=5.4cm,yshift=-9.5cm,scale=.7]
\coordinate (p1) at (-30:\rad);
\coordinate (p2) at (+30:\rad);
\coordinate (p3) at (+150:\rad);
\coordinate (p4) at (-150:\rad);
\coordinate (i1) at (0:\rad/3);
\coordinate (i2) at (180:\rad/3);
\draw [very thick,SeaGreen] ($(i1)!1/3!(i2)$) -- ++(-90:\rad/3);
\draw [very thick,SeaGreen] ($(i1)!2/3!(i2)$) -- ++(-90:\rad/3);
\draw [very thick,black] (p1) -- (i1) -- (p2);
\draw [very thick,black] (p3) -- (i2) -- (p4);
\draw [very thick,black] (i1) -- (i2);
\end{scope}
\begin{scope}[xshift=5.4cm,yshift=-10.75cm,scale=.7]
\coordinate (p1) at (-30:\rad);
\coordinate (p2) at (+30:\rad);
\coordinate (p3) at (+150:\rad);
\coordinate (p4) at (-150:\rad);
\coordinate (i1) at (0:\rad/3);
\coordinate (i2) at (180:\rad/3);
\draw [very thick,Mahogany] ($(i1)!1/3!(p1)$) -- ++(-132:\rad/3);
\draw [very thick,Mahogany] ($(i1)!2/3!(p1)$) -- ++(-132:\rad/3);
\draw [very thick,black] (p1) -- (i1) -- (p2);
\draw [very thick,black] (p3) -- (i2) -- (p4);
\draw [very thick,black] (i1) -- (i2);
\end{scope}
\begin{scope}[xshift=5.4cm,yshift=-12cm,scale=.7]
\coordinate (p1) at (-30:\rad);
\coordinate (p2) at (+30:\rad);
\coordinate (p3) at (+150:\rad);
\coordinate (p4) at (-150:\rad);
\coordinate (i1) at (0:\rad/3);
\coordinate (i2) at (180:\rad/3);
\draw [very thick,SeaGreen] ($(i1)!1/2!(p1)$) -- ++(-132:\rad/3);
\draw [very thick,SeaGreen] ($(i1)!1/2!(i2)$) -- ++(-90:\rad/3);
\draw [very thick,black] (p1) -- (i1) -- (p2);
\draw [very thick,black] (p3) -- (i2) -- (p4);
\draw [very thick,black] (i1) -- (i2);
\end{scope}
\end{tikzpicture}
\end{center}
The diagrams in the first two groups come with a minus sign and those in the last two a plus sign, thus all diagrams with the propagator $1/(k_++k_-)^2$ (
colored in magenta) cancel in pair. As expected, this cancellation corresponds to the cancellation of tadpoles on the one-loop side (due to anti-symmetry of the structure constants). The remaining diagrams are almost all those produced by inserting two new legs $+$ and $-$ into the original four-point tree from the bottom side avoiding the presence of $1/(k_++k_-)^2$:
\begin{equation}
\parbox{3cm}{\tikz{
\draw [thick,OrangeRed] (0,0) circle [radius=\rad];
\coordinate [label=-30:$1$] (p1) at (-30:\rad);
\coordinate [label=+30:$2$] (p2) at (+30:\rad);
\coordinate [label=+150:$3$] (p3) at (+150:\rad);
\coordinate [label=-150:$4$] (p4) at (-150:\rad);
\coordinate [label=-70:$-$] (minus) at (-70:\rad);
\coordinate [label=-110:$+$] (plus) at (-110:\rad);
\draw [very thick,Cyan] (p1) -- (p2) -- (p4) -- (p3) -- cycle;
\draw [very thick,Cyan,->>] (plus) -- (minus);
\draw [thick,Cyan,->>] (0:3*\rad/4) arc [start angle=0, end angle=340, radius=\rad/5];
\draw [thick,Cyan,->>] (180:3*\rad/4) arc [start angle=180, end angle=-160, radius=\rad/5]; 
\node [edot] at (p1) {};
\node [edot] at (p2) {};
\node [edot] at (p3) {};
\node [edot] at (p4) {};
\node [idot] at (minus) {};
\node [idot] at (plus) {};
\node [fill=Cyan,inner sep=2pt,label=90:$\color{Cyan}A$] at (0,0) {};
}}=
\parbox{1.5cm}{\tikz{
\begin{scope}[scale=.7]
\coordinate (p1) at (-30:\rad);
\coordinate (p2) at (+30:\rad);
\coordinate (p3) at (+150:\rad);
\coordinate (p4) at (-150:\rad);
\coordinate (i1) at (0:\rad/3);
\coordinate (i2) at (180:\rad/3);
\draw [very thick,Mahogany] ($(i2)!1/3!(p4)$) -- ++(-48:\rad/3);
\draw [very thick,Mahogany] ($(i2)!2/3!(p4)$) -- ++(-48:\rad/3);
\draw [very thick,black] (p1) -- (i1) -- (p2);
\draw [very thick,black] (p3) -- (i2) -- (p4);
\draw [very thick,black] (i1) -- (i2);
\end{scope}
}}+
\parbox{1.5cm}{\tikz{
\begin{scope}[scale=.7]
\coordinate (p1) at (-30:\rad);
\coordinate (p2) at (+30:\rad);
\coordinate (p3) at (+150:\rad);
\coordinate (p4) at (-150:\rad);
\coordinate (i1) at (0:\rad/3);
\coordinate (i2) at (180:\rad/3);
\draw [very thick,SeaGreen] ($(i1)!1/3!(i2)$) -- ++(-90:\rad/3);
\draw [very thick,SeaGreen] ($(i1)!2/3!(i2)$) -- ++(-90:\rad/3);
\draw [very thick,black] (p1) -- (i1) -- (p2);
\draw [very thick,black] (p3) -- (i2) -- (p4);
\draw [very thick,black] (i1) -- (i2);
\end{scope}
}}+
\parbox{1.5cm}{\tikz{
\begin{scope}[scale=.7]
\coordinate (p1) at (-30:\rad);
\coordinate (p2) at (+30:\rad);
\coordinate (p3) at (+150:\rad);
\coordinate (p4) at (-150:\rad);
\coordinate (i1) at (0:\rad/3);
\coordinate (i2) at (180:\rad/3);
\draw [very thick,Mahogany] ($(i1)!1/3!(p1)$) -- ++(-132:\rad/3);
\draw [very thick,Mahogany] ($(i1)!2/3!(p1)$) -- ++(-132:\rad/3);
\draw [very thick,black] (p1) -- (i1) -- (p2);
\draw [very thick,black] (p3) -- (i2) -- (p4);
\draw [very thick,black] (i1) -- (i2);
\end{scope}
}}+
\parbox{1.5cm}{\tikz{
\begin{scope}[scale=.7]
\coordinate (p1) at (-30:\rad);
\coordinate (p2) at (+30:\rad);
\coordinate (p3) at (+150:\rad);
\coordinate (p4) at (-150:\rad);
\coordinate (i1) at (0:\rad/3);
\coordinate (i2) at (180:\rad/3);
\draw [very thick,SeaGreen] ($(i1)!1/2!(p1)$) -- ++(-132:\rad/3);
\draw [very thick,SeaGreen] ($(i1)!1/2!(i2)$) -- ++(-90:\rad/3);
\draw [very thick,black] (p1) -- (i1) -- (p2);
\draw [very thick,black] (p3) -- (i2) -- (p4);
\draw [very thick,black] (i1) -- (i2);
\end{scope}
}}.
\end{equation}
There is an important subtlety here: on the RHS of the above equality, two diagrams that naively seem to be allowed, \tikz{\begin{scope}[scale=.25]
\coordinate (p1) at (-30:\rad);
\coordinate (p2) at (+30:\rad);
\coordinate (p3) at (+150:\rad);
\coordinate (p4) at (-150:\rad);
\coordinate (i1) at (0:\rad/3);
\coordinate (i2) at (180:\rad/3);
\draw [very thick,SeaGreen] ($(i2)!1/2!(p4)$) -- ++(-48:\rad/3);
\draw [very thick,SeaGreen] ($(i1)!1/2!(i2)$) -- ++(-90:\rad/3);
\draw [very thick,black] (p1) -- (i1) -- (p2);
\draw [very thick,black] (p3) -- (i2) -- (p4);
\draw [very thick,black] (i1) -- (i2);
\end{scope}} and \tikz{\begin{scope}[scale=.25]
\coordinate (p1) at (-30:\rad);
\coordinate (p2) at (+30:\rad);
\coordinate (p3) at (+150:\rad);
\coordinate (p4) at (-150:\rad);
\coordinate (i1) at (0:\rad/3);
\coordinate (i2) at (180:\rad/3);
\draw [very thick,SeaGreen] ($(i2)!1/2!(p4)$) -- ++(-48:\rad/3);
\draw [very thick,SeaGreen] ($(i1)!1/2!(p1)$) -- ++(-132:\rad/3);
\draw [very thick,black] (p1) -- (i1) -- (p2);
\draw [very thick,black] (p3) -- (i2) -- (p4);
\draw [very thick,black] (i1) -- (i2);
\end{scope}}, are actually absent. To understand why, note that the ordering $\rho=(1243)$ induces a specific ordering in each of the blue triangles, one clockwise and the other anti-clockwise. When we insert the pair $(+-)$ on the disk boundary it goes anti-clockwise. The absence of the two diagrams can be interpreted as following the rule that: when $+$ and $-$ are inserted into two different edges in the original tree diagram, the two edges must come from a single triangle whose ordering is the same as that of $(+-)$.

This result is very suggestive and can be immediately generalized. In the graphic representation for a generic $m_n^{\rm tree}[{\color{OrangeRed}\pi}|{\color{Cyan}\rho}]$, the pattern that $\rho$ forms is a set of \textit{ordered} polygons glued at their vertices. When we insert the two legs $+$ and $-$ from one side of the disk, we insert them in all possible ways to the tree structures that come from the polygon that has the same ordering as $(+-)$, but only to the same edges in the tree structures when the polygon has opposite ordering.

Armed with this diagrammatic rule, we are able to directly write out the result from the other three $(-\alpha+)$ orderings ({\it i.e.} inserting $(+-)$ on the other three sides)
\begin{align}
\parbox{3cm}{\tikz{
\draw [thick,OrangeRed] (0,0) circle [radius=\rad];
\coordinate [label=-30:$1$] (p1) at (-30:\rad);
\coordinate [label=+30:$2$] (p2) at (+30:\rad);
\coordinate [label=+150:$3$] (p3) at (+150:\rad);
\coordinate [label=-150:$4$] (p4) at (-150:\rad);
\coordinate [label=15:$-$] (minus) at (15:\rad);
\coordinate [label=-15:$+$] (plus) at (-15:\rad);
\draw [very thick,Cyan] (p1) -- (p2) -- (p4) -- (p3) -- cycle;
\draw [very thick,Cyan,->>] (plus) -- (minus);
\draw [thick,Cyan,->>] (0:3*\rad/4) arc [start angle=0, end angle=340, radius=\rad/5];
\draw [thick,Cyan,->>] (180:3*\rad/4) arc [start angle=180, end angle=-160, radius=\rad/5]; 
\node [edot] at (p1) {};
\node [edot] at (p2) {};
\node [edot] at (p3) {};
\node [edot] at (p4) {};
\node [idot] at (minus) {};
\node [idot] at (plus) {};
}}&=
\parbox{1.5cm}{\tikz{
\begin{scope}[scale=.7]
\coordinate (p1) at (-30:\rad);
\coordinate (p2) at (+30:\rad);
\coordinate (p3) at (+150:\rad);
\coordinate (p4) at (-150:\rad);
\coordinate (i1) at (0:\rad/3);
\coordinate (i2) at (180:\rad/3);
\draw [very thick,Mahogany] ($(i1)!1/3!(p1)$) -- ++(48:\rad/3);
\draw [very thick,Mahogany] ($(i1)!2/3!(p1)$) -- ++(48:\rad/3);
\draw [very thick,black] (p1) -- (i1) -- (p2);
\draw [very thick,black] (p3) -- (i2) -- (p4);
\draw [very thick,black] (i1) -- (i2);
\end{scope}
}}+
\parbox{1.5cm}{\tikz{
\begin{scope}[scale=.7]
\coordinate (p1) at (-30:\rad);
\coordinate (p2) at (+30:\rad);
\coordinate (p3) at (+150:\rad);
\coordinate (p4) at (-150:\rad);
\coordinate (i1) at (0:\rad/3);
\coordinate (i2) at (180:\rad/3);
\draw [very thick,Mahogany] ($(i1)!1/3!(p2)$) -- ++(-48:\rad/3);
\draw [very thick,Mahogany] ($(i1)!2/3!(p2)$) -- ++(-48:\rad/3);
\draw [very thick,black] (p1) -- (i1) -- (p2);
\draw [very thick,black] (p3) -- (i2) -- (p4);
\draw [very thick,black] (i1) -- (i2);
\end{scope}
}}+
\parbox{1.5cm}{\tikz{
\begin{scope}[scale=.7]
\coordinate (p1) at (-30:\rad);
\coordinate (p2) at (+30:\rad);
\coordinate (p3) at (+150:\rad);
\coordinate (p4) at (-150:\rad);
\coordinate (i1) at (0:\rad/3);
\coordinate (i2) at (180:\rad/3);
\draw [very thick,SeaGreen] ($(i1)!2/3!(p1)$) -- ++(+48:\rad/3);
\draw [very thick,SeaGreen] ($(i1)!2/3!(p2)$) -- ++(-48:\rad/3);
\draw [very thick,black] (p1) -- (i1) -- (p2);
\draw [very thick,black] (p3) -- (i2) -- (p4);
\draw [very thick,black] (i1) -- (i2);
\end{scope}
}}\;,\\
\parbox{3cm}{\tikz{
\draw [thick,OrangeRed] (0,0) circle [radius=\rad];
\coordinate [label=-30:$1$] (p1) at (-30:\rad);
\coordinate [label=+30:$2$] (p2) at (+30:\rad);
\coordinate [label=+150:$3$] (p3) at (+150:\rad);
\coordinate [label=-150:$4$] (p4) at (-150:\rad);
\coordinate [label=110:$-$] (minus) at (110:\rad);
\coordinate [label=70:$+$] (plus) at (70:\rad);
\draw [very thick,Cyan] (p1) -- (p2) -- (p4) -- (p3) -- cycle;
\draw [very thick,Cyan,->>] (plus) -- (minus);
\draw [thick,Cyan,->>] (0:3*\rad/4) arc [start angle=0, end angle=340, radius=\rad/5];
\draw [thick,Cyan,->>] (180:3*\rad/4) arc [start angle=180, end angle=-160, radius=\rad/5]; 
\node [edot] at (p1) {};
\node [edot] at (p2) {};
\node [edot] at (p3) {};
\node [edot] at (p4) {};
\node [idot] at (minus) {};
\node [idot] at (plus) {};
}}&=
\parbox{1.5cm}{\tikz{
\begin{scope}[scale=.7]
\coordinate (p1) at (-30:\rad);
\coordinate (p2) at (+30:\rad);
\coordinate (p3) at (+150:\rad);
\coordinate (p4) at (-150:\rad);
\coordinate (i1) at (0:\rad/3);
\coordinate (i2) at (180:\rad/3);
\draw [very thick,Mahogany] ($(i1)!1/3!(p2)$) -- ++(132:\rad/3);
\draw [very thick,Mahogany] ($(i1)!2/3!(p2)$) -- ++(132:\rad/3);
\draw [very thick,black] (p1) -- (i1) -- (p2);
\draw [very thick,black] (p3) -- (i2) -- (p4);
\draw [very thick,black] (i1) -- (i2);
\end{scope}
}}+
\parbox{1.5cm}{\tikz{
\begin{scope}[scale=.7]
\coordinate (p1) at (-30:\rad);
\coordinate (p2) at (+30:\rad);
\coordinate (p3) at (+150:\rad);
\coordinate (p4) at (-150:\rad);
\coordinate (i1) at (0:\rad/3);
\coordinate (i2) at (180:\rad/3);
\draw [very thick,SeaGreen] ($(i1)!1/3!(i2)$) -- ++(90:\rad/3);
\draw [very thick,SeaGreen] ($(i1)!2/3!(i2)$) -- ++(90:\rad/3);
\draw [very thick,black] (p1) -- (i1) -- (p2);
\draw [very thick,black] (p3) -- (i2) -- (p4);
\draw [very thick,black] (i1) -- (i2);
\end{scope}
}}+
\parbox{1.5cm}{\tikz{
\begin{scope}[scale=.7]
\coordinate (p1) at (-30:\rad);
\coordinate (p2) at (+30:\rad);
\coordinate (p3) at (+150:\rad);
\coordinate (p4) at (-150:\rad);
\coordinate (i1) at (0:\rad/3);
\coordinate (i2) at (180:\rad/3);
\draw [very thick,Mahogany] ($(i2)!1/3!(p3)$) -- ++(48:\rad/3);
\draw [very thick,Mahogany] ($(i2)!2/3!(p3)$) -- ++(48:\rad/3);
\draw [very thick,black] (p1) -- (i1) -- (p2);
\draw [very thick,black] (p3) -- (i2) -- (p4);
\draw [very thick,black] (i1) -- (i2);
\end{scope}
}}+
\parbox{1.5cm}{\tikz{
\begin{scope}[scale=.7]
\coordinate (p1) at (-30:\rad);
\coordinate (p2) at (+30:\rad);
\coordinate (p3) at (+150:\rad);
\coordinate (p4) at (-150:\rad);
\coordinate (i1) at (0:\rad/3);
\coordinate (i2) at (180:\rad/3);
\draw [very thick,SeaGreen] ($(i1)!1/2!(p2)$) -- ++(132:\rad/3);
\draw [very thick,SeaGreen] ($(i1)!1/2!(i2)$) -- ++(90:\rad/3);
\draw [very thick,black] (p1) -- (i1) -- (p2);
\draw [very thick,black] (p3) -- (i2) -- (p4);
\draw [very thick,black] (i1) -- (i2);
\end{scope}
}}\;,\\
\parbox{3cm}{\tikz{
\draw [thick,OrangeRed] (0,0) circle [radius=\rad];
\coordinate [label=-30:$1$] (p1) at (-30:\rad);
\coordinate [label=+30:$2$] (p2) at (+30:\rad);
\coordinate [label=+150:$3$] (p3) at (+150:\rad);
\coordinate [label=-150:$4$] (p4) at (-150:\rad);
\coordinate [label=-165:$-$] (minus) at (-165:\rad);
\coordinate [label=165:$+$] (plus) at (165:\rad);
\draw [very thick,Cyan] (p1) -- (p2) -- (p4) -- (p3) -- cycle;
\draw [very thick,Cyan,->>] (plus) -- (minus);
\draw [thick,Cyan,->>] (0:3*\rad/4) arc [start angle=0, end angle=340, radius=\rad/5];
\draw [thick,Cyan,->>] (180:3*\rad/4) arc [start angle=180, end angle=-160, radius=\rad/5]; 
\node [edot] at (p1) {};
\node [edot] at (p2) {};
\node [edot] at (p3) {};
\node [edot] at (p4) {};
\node [idot] at (minus) {};
\node [idot] at (plus) {};
}}&=
\parbox{1.5cm}{\tikz{
\begin{scope}[scale=.7]
\coordinate (p1) at (-30:\rad);
\coordinate (p2) at (+30:\rad);
\coordinate (p3) at (+150:\rad);
\coordinate (p4) at (-150:\rad);
\coordinate (i1) at (0:\rad/3);
\coordinate (i2) at (180:\rad/3);
\draw [very thick,Mahogany] ($(i2)!1/3!(p3)$) -- ++(-132:\rad/3);
\draw [very thick,Mahogany] ($(i2)!2/3!(p3)$) -- ++(-132:\rad/3);
\draw [very thick,black] (p1) -- (i1) -- (p2);
\draw [very thick,black] (p3) -- (i2) -- (p4);
\draw [very thick,black] (i1) -- (i2);
\end{scope}
}}+
\parbox{1.5cm}{\tikz{
\begin{scope}[scale=.7]
\coordinate (p1) at (-30:\rad);
\coordinate (p2) at (+30:\rad);
\coordinate (p3) at (+150:\rad);
\coordinate (p4) at (-150:\rad);
\coordinate (i1) at (0:\rad/3);
\coordinate (i2) at (180:\rad/3);
\draw [very thick,Mahogany] ($(i2)!1/3!(p4)$) -- ++(132:\rad/3);
\draw [very thick,Mahogany] ($(i2)!2/3!(p4)$) -- ++(132:\rad/3);
\draw [very thick,black] (p1) -- (i1) -- (p2);
\draw [very thick,black] (p3) -- (i2) -- (p4);
\draw [very thick,black] (i1) -- (i2);
\end{scope}
}}\;.
\end{align}
We obtain the RHS of \eqref{example1} by summing over the above four results and taking the forward limit. An important fact is that diagrams with both $+$ and $-$ inserted onto a single external leg in the original tree (colored in brown) always come in pairs, and we have the relation
\begin{equation}
\lim_{k_\pm\to\pm\ell}\left(
\parbox{1.5cm}{\tikz{
\begin{scope}[scale=.7]
\coordinate (p1) at (0:\rad/3);
\coordinate [label=0:$a$] (p2) at (0:\rad);
\draw [very thick,Mahogany] ($(p1)!1/3!(p2)$) -- ++(-90:\rad/3);
\draw [very thick,Mahogany] ($(p1)!2/3!(p2)$) -- ++(-90:\rad/3);
\draw [very thick,black] (0,0) -- (p2);
\node [circle,draw=black,thick,fill=gray,inner sep=\rad/7] at (0,0) {};
\end{scope}
}}+
\parbox{1.5cm}{\tikz{
\begin{scope}[scale=.7]
\coordinate (p1) at (0:\rad/3);
\coordinate [label=0:$a$] (p2) at (0:\rad);
\draw [very thick,Mahogany] ($(p1)!1/3!(p2)$) -- ++(90:\rad/3);
\draw [very thick,Mahogany] ($(p1)!2/3!(p2)$) -- ++(90:\rad/3);
\draw [very thick,black] (0,0) -- (p2);
\node [circle,draw=black,thick,fill=gray,inner sep=\rad/7] at (0,0) {};
\end{scope}
}}\right)=
-\frac{1}{4(l\cdot k_a)^2}\,\times
\parbox{1.5cm}{\tikz{
\begin{scope}[scale=.7]
\coordinate (p1) at (0:\rad/3);
\coordinate [label=0:$a$] (p2) at (0:\rad);
\draw [very thick,black] (0,0) -- (p2);
\node [circle,draw=black,thick,fill=gray,inner sep=\rad/7] at (0,0) {};
\end{scope}
}}\,,\quad\forall a\,,
\end{equation}
where the gray blob denotes the remaining part of the original tree. They yield the diagrams with external-leg bubbles, which after loop integration add up to the tree amplitude multiplied by the overall integral $\Delta_4$, which vanishes in $D>6$. 

For the remaining diagrams (colored in green), it is not hard to observe that they separate into groups, where each group matches the pattern of the RHS of \eqref{looptreegraph}. Hence after taking the forward limit they form a triangle and a bubble, which respect ordering $(1234)$ and $(1243)$,  in the new representation~\eqref{newrep}. 
This is exactly what we expect from \eqref{1loopFD} (up to the vanishing terms after the loop integration), thus proving \eqref{example1}. Explicitly we have:
\begin{equation}
m_4^{1\text{-loop}}[1234|1243]=
\parbox{2.5cm}{\tikz{
\begin{scope}[scale=.8]
\draw [very thick,black] (-30:\rad) -- (0:\rad/2) -- (30:\rad);
\draw [very thick,black] (-150:\rad) -- (180:\rad/2) -- (150:\rad);
\draw [very thick,black] (0:\rad/2) -- (180:\rad/2);
\node [circle,very thick,draw=black,fill=white,inner sep=\rad/7] at (0:\rad/2) {};
\draw [very thick,black,->] (0:\rad/2) ++(10:.24*\rad) arc [start angle=10,end angle=370, radius=.24*\rad];
\coordinate [label=-30:$1$] (p1) at (-30:\rad);
\coordinate [label=+30:$2$] (p2) at (+30:\rad);
\coordinate [label=+150:$3$] (p3) at (+150:\rad);
\coordinate [label=-150:$4$] (p4) at (-150:\rad);
\end{scope}
}}+
\parbox{2.5cm}{\tikz{
\begin{scope}[scale=.8]
\draw [very thick,black] (-30:\rad) -- (0:\rad/2) -- (30:\rad);
\draw [very thick,black] (-150:\rad) -- (180:\rad/2) -- (150:\rad);
\draw [very thick,black] (0:\rad/2) -- (180:\rad/2);
\node [circle,very thick,draw=black,fill=white,inner sep=\rad/7] at (0,0) {};
\draw [very thick,black,->] (100:.24*\rad) arc [start angle=100,end angle=460, radius=.24*\rad];
\coordinate [label=-30:$1$] (p1) at (-30:\rad);
\coordinate [label=+30:$2$] (p2) at (+30:\rad);
\coordinate [label=+150:$3$] (p3) at (+150:\rad);
\coordinate [label=-150:$4$] (p4) at (-150:\rad);
\end{scope}
}}+\Delta_4\,m_4^{\rm tree}[1234|1243]\;.
\end{equation}

Next we study another example where the two orderings are the same:
\begin{equation}\label{example2}
m_4^{1\text{-loop}}[1234|1234]=\int\frac{d^D \ell}{(2\pi)^D}\frac{1}{\ell^2}\lim_{k_\pm\to\pm\ell}\sum_{\alpha,\beta\in\text{cyc}(1234)}\hspace{-.5em}m_6^{\rm tree}[-\alpha+|-\beta+]
\end{equation}
and the starting point is a graph in which the blue loop forms a square\begin{equation}\label{example2tree}
\parbox{3cm}{\tikz{
\begin{scope}
\draw [thick,OrangeRed] (0,0) circle [radius=\rad];
\coordinate [label=-45:$1$] (p1) at (-45:\rad);
\coordinate [label=+45:$2$] (p2) at (+45:\rad);
\coordinate [label=+135:$3$] (p3) at (+135:\rad);
\coordinate [label=-135:$4$] (p4) at (-135:\rad);
\draw [very thick,Cyan] (p1) -- (p2) -- (p3) -- (p4) -- cycle;
\draw [very thick,Cyan,->>] (0:\rad/2) arc [start angle=0,end angle=340,radius=\rad/2];
\node [edot] at (p1) {};
\node [edot] at (p2) {};
\node [edot] at (p3) {};
\node [edot] at (p4) {};
\end{scope}
}}=-
\parbox{2.5cm}{\tikz{
\begin{scope}[scale=.8]
\draw [very thick,black] (-45:\rad) -- (0:\rad/3) -- (45:\rad);
\draw [very thick,black] (-135:\rad) -- (180:\rad/3) -- (135:\rad);
\draw [very thick,black] (0:\rad/3) -- (180:\rad/3);
\coordinate [label=-45:$1$] (p1) at (-45:\rad);
\coordinate [label=+45:$2$] (p2) at (+45:\rad);
\coordinate [label=+135:$3$] (p3) at (+135:\rad);
\coordinate [label=-135:$4$] (p4) at (-135:\rad);
\end{scope}
}}-
\parbox{2.5cm}{\tikz{
\begin{scope}[scale=.8]
\draw [very thick,black] (45:\rad) -- (90:\rad/3) -- (135:\rad);
\draw [very thick,black] (-45:\rad) -- (-90:\rad/3) -- (-135:\rad);
\draw [very thick,black] (90:\rad/3) -- (-90:\rad/3);
\coordinate [label=-45:$1$] (p1) at (-45:\rad);
\coordinate [label=+45:$2$] (p2) at (+45:\rad);
\coordinate [label=+135:$3$] (p3) at (+135:\rad);
\coordinate [label=-135:$4$] (p4) at (-135:\rad);
\end{scope}
}}=-\frac{1}{(k_1+k_2)^2}-\frac{1}{(k_2+k_3)^2}\,.
\end{equation}
We can do the explicit calculations as in the previous case, but let us try the 
rule summarized from that case.

When we insert the $(+-)$ pair, the ordering is the same as that of the blue box, so we should insert $+$ and $-$ in all possible ways into the two tree diagrams, such that the two labels are place on the same side with the correct ordering and that the pole $1/(k_++k_-)^2$ is absent. The final result is exactly one box, four triangles, two bubbles, plus the $\Delta_4$ times tree:
\begin{equation}\label{example2result1}
\begin{split}
m_4^{1\text{-loop}}[1234|1234]=&-
\parbox{2.5cm}{\tikz{
\begin{scope}[scale=.8]
\draw [very thick,black] (-45:\rad) -- (-45:\rad/2) -- (45:\rad/2) -- (45:\rad);
\draw [very thick,black] (-135:\rad) -- (-135:\rad/2) -- (135:\rad/2) -- (135:\rad);
\draw [very thick,black] (-45:\rad/2) -- (-135:\rad/2);
\draw [very thick,black] (45:\rad/2) -- (135:\rad/2);
\draw [very thick,black,->] (-45:\rad/2) -- ($(-45:\rad/2)!.6!(45:\rad/2)$);
\coordinate [label=-45:$1$] (p1) at (-45:\rad);
\coordinate [label=+45:$2$] (p2) at (+45:\rad);
\coordinate [label=+135:$3$] (p3) at (+135:\rad);
\coordinate [label=-135:$4$] (p4) at (-135:\rad);
\end{scope}
}}-\left(
\parbox{2.5cm}{\tikz{
\begin{scope}[scale=.8]
\draw [very thick,black] (-30:\rad) -- (0:\rad/2) -- (30:\rad);
\draw [very thick,black] (-150:\rad) -- (180:\rad/2) -- (150:\rad);
\draw [very thick,black] (0:\rad/2) -- (180:\rad/2);
\node [circle,very thick,draw=black,fill=white,inner sep=\rad/7] at (0:\rad/2) {};
\draw [very thick,black,->] (0:\rad/2) ++(10:.24*\rad) arc [start angle=10,end angle=370, radius=.24*\rad];
\coordinate [label=-30:$1$] (p1) at (-30:\rad);
\coordinate [label=+30:$2$] (p2) at (+30:\rad);
\coordinate [label=+150:$3$] (p3) at (+150:\rad);
\coordinate [label=-150:$4$] (p4) at (-150:\rad);
\end{scope}
}}+\text{cyc}\right)\\&-\left(
\parbox{2.5cm}{\tikz{
\begin{scope}[scale=.8]
\draw [very thick,black] (-30:\rad) -- (0:\rad/2) -- (30:\rad);
\draw [very thick,black] (-150:\rad) -- (180:\rad/2) -- (150:\rad);
\draw [very thick,black] (0:\rad/2) -- (180:\rad/2);
\node [circle,very thick,draw=black,fill=white,inner sep=\rad/7] at (0,0) {};
\draw [very thick,black,->] (100:.24*\rad) arc [start angle=100,end angle=460, radius=.24*\rad];
\coordinate [label=-30:$1$] (p1) at (-30:\rad);
\coordinate [label=+30:$2$] (p2) at (+30:\rad);
\coordinate [label=+150:$3$] (p3) at (+150:\rad);
\coordinate [label=-150:$4$] (p4) at (-150:\rad);
\end{scope}
}}+\text{cyc}\right)+\Delta_4\,m_4^{\rm tree}[1234|1234]\,.
\end{split}
\end{equation}
Alternatively, if we consider $m_4^{1\text{-loop}}[1234|4321]$, although the corresponding four-point tree amplitude is the same, the ordering of the blue box in \eqref{example2tree} becomes opposite, and so the leg $+$ and the leg $-$ are only allowed to insert onto the same edge in the original tree diagrams. As a consequence the box and the triangles are absent and
\begin{equation}\label{example2result2}
\begin{split}
m_4^{1\text{-loop}}[1234|4321]=
-\left(
\parbox{2.5cm}{\tikz{
\begin{scope}[scale=.8]
\draw [very thick,black] (-30:\rad) -- (0:\rad/2) -- (30:\rad);
\draw [very thick,black] (-150:\rad) -- (180:\rad/2) -- (150:\rad);
\draw [very thick,black] (0:\rad/2) -- (180:\rad/2);
\node [circle,very thick,draw=black,fill=white,inner sep=\rad/7] at (0,0) {};
\draw [very thick,black,->] (100:.24*\rad) arc [start angle=100,end angle=460, radius=.24*\rad];
\coordinate [label=-30:$1$] (p1) at (-30:\rad);
\coordinate [label=+30:$2$] (p2) at (+30:\rad);
\coordinate [label=+150:$3$] (p3) at (+150:\rad);
\coordinate [label=-150:$4$] (p4) at (-150:\rad);
\end{scope}
}}+\text{cyc}\right)+\Delta_4\,m_4^{\rm tree}[1234|4321]\,.
\end{split}
\end{equation}
The results \eqref{example2result1} and \eqref{example2result2} are explicitly checked, which confirms our diagrammatic rules acquired from the computations in verifying \eqref{example1}.

\section{Conclusion and discussions}\label{sec:discussion}

In this paper we showed that one-loop scattering equations, the $\SL2C$-covariant form of the equations derived in \cite{Geyer:2015bja}, are identical to the tree-level scattering equations involving two additional massive particles in the forward limit. Properties of the solutions to the scattering equations, both at tree and one-loop level, are studied from this point of view, and the prescription for the general CHY formulation at one loop is presented.

The connection between one-loop amplitudes and forward limit of trees with two massive particles is of interests regardless of any formulation of amplitudes. As we have shown in the bi-adjoint scalar theory, one-loop double-partial amplitudes can be obtained from such forward limit of tree-level double-partial amplitudes, \eqref{partialamp}. The result is in the new representation of one-loop integrals introduced in~\cite{Geyer:2015bja}, and we have seen that it can now be explained as originated from massive tree amplitudes in forward limit. Recall that \eqref{partialamp} is equivalent to the relations between color-dressed amplitudes, thus \eqref{question} is indeed an identity. Given that CHY formulas for tree amplitudes is known, this connection between one loop and tree directly leads to a closed formula for the one-loop amplitudes based on the one-loop scattering equations, as we presented in \eqref{1loopphi3} (or the corresponding color-dressed version). 

We strongly suspect that \eqref{question} may be valid for more general theories. A direct manipulation on Feynman diagrams, as we have done for bi-adjoint scalars, may prove difficult for more sophisticated theories. However, based on CHY formula or other useful formulation, one can take forward limit of other theories with two massive scalars, and it would be highly desirable to see if the resulting formula again gives correct one-loop amplitudes. Taking forward limit of particles with non-trivial spins is more subtle: one has to start with the appropriate massive tree amplitudes, and perform the sum over internal states of the massive particles. We leave these very intriguing questions to future works. 

In another aspect, whenever a connection between one-loop amplitudes and tree amplitudes via such forward limit is confirmed, a CHY formula for the one-loop amplitudes should in principle be induced from the corresponding tree-level ones. A subtlety is that, in order this works one has to ensure that in the forward limit the formula on the tree-level side receives vanishing contributions from the solutions that are excluded at one-loop (\textit{i.e.} the singular II sector). It is interesting to note that such criteria may not be met by every CHY formula for the same tree amplitude. As an example, inspired by \eqref{looptreegraph} we have a CHY formula for a single triangle diagram as follows
\begin{equation}
\begin{split}
\parbox{2.5cm}{\tikz{
\begin{scope}[scale=.8]
\draw [very thick,black] (-30:\rad) -- (0:\rad/2) -- (30:\rad);
\draw [very thick,black] (-150:\rad) -- (180:\rad/2) -- (150:\rad);
\draw [very thick,black] (0:\rad/2) -- (180:\rad/2);
\node [circle,very thick,draw=black,fill=white,inner sep=\rad/7] at (0:\rad/2) {};
\draw [very thick,black,->] (0:\rad/2) ++(10:.24*\rad) arc [start angle=10,end angle=370, radius=.24*\rad];
\coordinate [label=-30:$1$] (p1) at (-30:\rad);
\coordinate [label=+30:$2$] (p2) at (+30:\rad);
\coordinate [label=+150:$3$] (p3) at (+150:\rad);
\coordinate [label=-150:$4$] (p4) at (-150:\rad);
\end{scope}
}}=
\int&\frac{d^D\ell}{(2\pi)^D}\frac{1}{\ell^2}\int\mu_4^{(1)}\Big(PT(1+2-34)PT(342-1+)\\[-.5em]
&-PT(12+34-)PT(341+2-)-PT(1-234+)PT(34+12-)\Big)\,,
\end{split}
\end{equation}
where each term in the integrand corresponds to each tree diagram on the RHS of \eqref{looptreegraph}. One may try to replace the first term by $PT(1+-234)(34-21+)$ since on the tree-level side it produces the same diagram before the forward limit, but this new integrand yields different result on the one-loop side.

\textit{Acknowledgments:} We thank Freddy Cachazo for many useful discussions. Research at Perimeter Institute is supported by the Government of Canada through Industry Canada and by the Province of Ontario through the Ministry of Research \& Innovation.

\bibliographystyle{apsrev4-1}
\bibliography{forward_limit}

\end{document}